\documentclass[aps,amsmath,amssymb,amsfonts,twocolumn,pra,showpacs]{revtex4-1}
\usepackage{graphicx}
\usepackage{bm}
\usepackage{color}
\usepackage{amsmath,amsthm,amssymb}
\usepackage{subfigure}

\usepackage{epstopdf}

\newcommand{\mrm}{\mathrm}

\begin{document}
\author{Micha{\l} Krych}
\affiliation{Faculty of Physics, University of Warsaw, Pasteura 5, 02-093 Warsaw, Poland}
\author{Zbigniew Idziaszek}
\affiliation{Faculty of Physics, University of Warsaw, Pasteura 5, 02-093 Warsaw, Poland}
\title{Reactive collisions of polar molecules in quasi-two-dimensional traps}

\begin{abstract}
We investigate collisions of polar molecules in quasi-2D traps in the presence of an external electric field perpendicular to the collision plane. We use the quantum-defect model characterized by two dimensionless parameters: $y$ and $s$. The first of them is related to the probability of the reaction at short distances, whereas the latter one defines the wave function phase at short distances. For $y$ close to unity we obtain universal collision rates determined by the quantum reflection process from the long-range part of the interaction potential that depends only on the van der Waals coefficient, dipole-dipole interaction and the trap frequency. For small short-range reaction probabilities collision rates are highly nonuniversal and trap induced shape resonances are visible. For high dipole moments we observe the damping of reactive collisions, which can stabilize the ultracold gas of polar molecules. The calculations were performed with help of multichannel wave funcion propagation by imposing short-range boundary condition derived from the quantum-defect model. 
\end{abstract}
\pacs{34.50.Cx,34.50.Lf}

\maketitle

\section{Introduction}

Successful realization of relatively dense ultracold gases of molecules in traps~\cite{Ni2008,Danzl2008,Molony2014,Takehoshi2014,Park2015} and optical lattices~\cite{Rb2ground,Danzl2010,Covey2015} in their rovibrational ground state offers new possibilities for physics and chemistry~\cite{Doyle2004,Carr2009,PSJ2012}. Ultracold molecule reaction rates can be relatively large as it was recently demonstrated for fermionic $^{40}$K$^{87}$Rb~\cite{Ospelkaus2010,Ni2010}. This highly reactive molecule belongs to the class of molecules, that have universal elastic and reactive collision rates. They can be calculated analytically from the knowledge of the long range part of the interaction potential only, provided the reaction probability at small distances is equal to unity \cite{Idziaszek2010,Quemener2010}. When reaction probability at short distances is smaller than one, the collision rates become dependent on the short range phase and the effect of resonances have to be taken into account \cite{Idziaszek2010b,Jachymski2013,Jachymski2014}.

Application of an external electric field to a gas of rotationless polar molecules polarizes them, and as a consequence increases the rates of reactive collisions leading to significant reduction of dipolar gas lifetime~\cite{Ni2010,Quemener2010,Idziaszek2010b}. The essential condition for creation of stable and dense samples of reactive ultracold molecules is the ability to understand and control of the loss process.
It was predicted~\cite{Buchler2007}, that reactive collisions of polar molecules can be stabilized by confining them in a quasi-two-dimensional (quasi-2D) geometry with the dipoles aligned perpendicular to the 2D plane. Reduction of reactive collisions is explained by the fact, that the molecules are ordered in a way, that they repeal each other, hence the probability of a short-range collision is significantly reduced. This reduction has been confirmed by theoretical calculations based on adiabatic potentials~\cite{Ticknor2010}, multichannel quantum dynamics for fermionic KRb~\cite{Quemener2010b,Micheli2010,Quemener2011} and by analytical models for
universal polar molecules with unit reaction probability at short distances~\cite{Micheli2010}. A similar effect can be achieved by confining polar molecules in a quasi-one-dimensional (quasi-1D) trap with dipole moments oriented in the direction perpendicular to the trap axis. This however does not lead to any further reduction of reaction rates in comparison to quasi-2D case \cite{simoni2015}.

In this paper we characterize elastic and reactive collisions in quasi-2D traps both for universal and nonuniversal polar molecules. The latter have probability of a short-range reaction smaller than one, which is described by a single dimensionless parameter. Hence, our results are applicable to a broad range of polar molecules. We investigate how the collision rates change with tuning of the characteristic trap length, dipolar length and the collision energy. Considering heteronuclear bialkali polar molecules that are at the heart of experimental investigations,  KRb, LiNa, LiK, LiRb and LiCs have allowed reaction channels ~\cite{Zuchowski2010}, henceforth their collisions should be universal. On the other hand, NaK, NaRb, NaCs, KCs, and RbCs do not have reactive channels, so they are expected to exhibit nonuniversal collisions in their ground state. In general, even nonuniversal polar molecules or homonuclear dimers can exhibit universal properties when prepared in an excited rotational or vibrational state~\cite{Julienne2009,Hudson2008}. While universal molecules do not have scattering resonances, because their close to threshold bound states vanish too fast, nonuniversal systems should exhibit dense spectrum of overlapping resonances. The latter case can be treated by models combing quantum-defect methods with assumptions derived from random-matrix theory \cite{Bohn2012,Mayle2013}.

The control of reactive collisions has been recently demonstrated for fermionic $^{40}$K$^{87}$Rb molecule in the rovibrational ground state~\cite{Ospelkaus2010,Ni2010,Miranda2011}. In the absence of an external electric field, molecular interaction is dominated by isotropic van der Waals potential, and ultracold fermionic molecules can either collide is $p$ or $s$ wave depending if they are prepared in identical or nonidentical internal spin state, respectively. As the ultracold $p$-wave collisions require tunnelling through a centrifugal barrier, the identical fermions reactions are significantly reduced at ultralow temperatures in comparison to distinguishable particles~\cite{Ospelkaus2010}. For a sample polarized by some external electric field, the reaction rates in three dimensional traps were increasing with the value of the induced dipole moment \cite{Ni2010}. Confining molecules in 1D deep optical lattice creates an effective array of quasi-2D traps~\cite{Miranda2011}. In this case polarizing the molecules in the direction parallel to the axis of quasi-2D traps, reduces the reactive collisions in accordance with the theoretical models, provided the molecules were prepared in the identical spin states~\cite{Miranda2011}. On the contrary, for collisions in two distinct spin states when the $s$-wave collision channel is available, inclusion of an external electric field only increases the reactive rates similarly to three-dimensional case.

This work is structured as follows. In section II we present our theoretical model that is used to describe reactive collisions in quasi-2D traps. In particular, we discuss the limit of a zero-dipole moment when the collision rates can be determined analytically from the properties of the van der Waals potential. In section III we present numerical results based on mutichannel wave function propagation. We discuss dependence of the collision rates on the dipolar length, confinement length and the kinetic energy. Finally we conclude in Sec. IV.

\section{Theoretical background}

\begin{figure}
  \begin{center}
    \includegraphics[width=0.7\linewidth,clip]{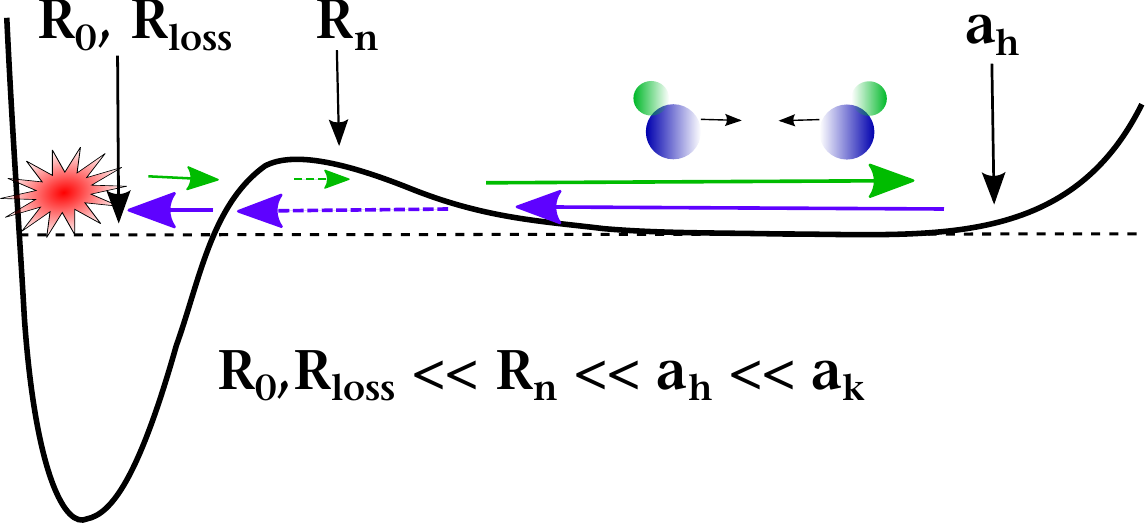}
  \end{center}
  \caption{Lengthscales separation in two polar molecules collision. $R_0$, $R_{\mrm{loss}}$ - the lengthscale of chemical reactions (losses), $R_n$ - van der Waals length, $a_h$ - characteristic length of the trap, $a_k$ - de Broglie wavelength.}
  \label{fig:separacja}
\end{figure}

Interaction of polar molecules differs significantly from the interaction of neutral atoms. In case of neutral atoms short-range interaction dominates and isotropic van der Waals potential is characterized by a $C_6$ coefficient.
On the other hand polar molecules interact via a long-range and anisotropic potential (it depends on the orientation of molecules in space)
\begin{equation}
V(r)=\frac{d(F)^2}{r^3}(1-3\cos^2\theta)-\frac{C_6}{r^6},
\end{equation}
where $d(F)$ is a dipole moment induced by an external electric field $F$ and $\theta$ is an angle between dipoles.
In order to calculate the collisonal dynamics of polar molecules we have used quantum defect method \cite{Jachymski2013, Jachymski2014}. It assumes the separation of lengthscales in the described problem, that is visualized in Fig.~\ref{fig:separacja}.

Elastic collision rates ($\mathcal{K}^\mrm{el}_{\ell m}$ and $ \mathcal{K}^\mrm{re}_{\ell m}$, respectively) can be described by a complex scattering length~\cite{Idziaszek2010}, which is directly connected to the scattering matrix~$S$
\begin{align}
\label{rozpraszania}
\tilde{a}_{\ell m}(E) & = \tilde{\alpha}_{\ell m}(E) - i \tilde{\beta}_{\ell m}(E) =
\frac{1}{i k} \frac{1-S_{\ell m,\ell m}}{1+S_{\ell m,\ell m}},
\end{align}
where $S_{\ell m,\ell m}$ are diagonal elements of the scattering matrix, $\ell m$ are angular-momentum quantum numbers, $E$ is a collision energy. Henceforth collision rates are given by
\begin{align}
\label{szybkosci}
\mathcal{K}^\mrm{el}_{\ell m}(E) & = g \frac{\pi \hbar}{\mu k} \left| 1 - S_{\ell m,\ell m}(E) \right|^2 = 2 g \frac{h k}{\mu} |\tilde{a}_{\ell m}(k)|^2 f_{\ell m}(k) \,,   \\
\label{szybkosci2}
 \mathcal{K}^\mrm{re}_{\ell m}(E) & = g \frac{\pi \hbar}{\mu k} \left(1- |S_{\ell m,\ell m}(E)|^2\right)  = 2g \frac{h}{\mu} \tilde{\beta}_{\ell m}(k) f_{\ell m}(k) \,.
\end{align}
where $\mu$ is a reduced mass, $k$ denote the wave vector, $g=1$ besides the case, when two particles are identical ($g=2$). Angular momentum is restricted to even numbers in case of identical bosons and to odd numbers in case of fermions in an identical state. Moreover $f_{\ell m}(k)$ is given by a formula
\begin{equation}
f_{\ell m}(k) = \frac{1}{1+k^2|\tilde{a}_{\ell m}(k)|^2+2k \tilde{\beta}_{\ell m}(k)}.
\end{equation}
For low energies $f_{\ell m}(k) \to 1$, when $k \to 0$, given that $k |\tilde{a}_{\ell m}(k)| \ll 1$ and $k \beta_{\ell m}(k) \ll 1$. In general $0 < f_{\ell m}(k) \leq 1$. On the ground of above equations \eqref{szybkosci} and \eqref{szybkosci2} one can state that at ultralow temperatures imaginary part of the scattering length is responsible for reactive scattering rates and modulus of the scattering length is responsible for elastic scattering rates.
%
%

We note that for small molecule separations wave functions can be written in a WKB like form
\begin{equation}
\begin{split}
\Psi(r)\propto & \frac{\exp[-i\int^r k(x) dx-i\varphi]}{\sqrt{(k(r))}}  \\
& - \frac{1-y}{1+y}\frac{\exp[i\int^r k(x) dx+i \varphi]}{\sqrt{(k(r))}},
\label{Psi}
\end{split}
\end{equation}
where $y$ gives an amplitude of the reflected wave function. For $y=1$ there is no reflected probability flux and short range reaction has a unit probability. Whereas for $y=0$ all incoming particles are reflected. At short distances van der Waals interaction is dominating, so we can use well known solutions for the zero energy radial wave function in a power-law potential~\cite{Sadeghpour2000}
\begin{equation}
\sqrt{r}R_{\ell,E=0}\propto J_{2\ell+1/4}\left(\frac{R_6^2}{2r^2}\right)+\tan(\varphi)Y_{2\ell+1/4}\left(\frac{R_6^2}{2r^2}\right),
\end{equation}
where $R_6 = \left(2 \mu C_6/ \hbar^2\right)^{1/4}$. Hence dimensionless scattering length $s=a/\bar{a}$ is connected to the short range phase by
\begin{equation}
s=\sqrt{2}\frac{\cos(\varphi-\pi/8)}{\sin(\varphi+\pi/8)},
\end{equation}
where we have used the definition of the average scattering length~\cite{Gribakin1993}
\begin{equation}
\bar{a}= \frac{2 \pi}{\Gamma\left(\frac14\right)^2}R_6.
\end{equation}
\begin{table}
\begin{center}
\begin{tabular}{|c|c|c|c|}
  \hline\hline
  N & 3D & 2D & 1D \\
  \hline
  $g_0$ & $1/\pi$ &$2/\pi$ & $2$ \\
  $g_1$ & $1/\pi$ &$4/\pi$ & $6$ \\
  \hline \hline
\end{tabular}
\begin{tabular}{|c|c|c|c|}
  \hline\hline
  N & 3D & 2D & 1D \\
  \hline
  $L_0$ & $\bar{a}$ & $\sqrt{\pi}\bar{a}$ & $2\bar{a}$ \\
  $L_1$ & $(k\bar{a})^2 \bar{a}_1$ & $(3\sqrt{\pi}/2)(q \bar{a})^2 \bar{a}_1$ & $6(p\bar{a})^2\bar{a}_1$ \\
  \hline \hline
\end{tabular}
\caption{Coefficients present in the formula for universal reactive scattering rate \eqref{uniw} for different dimensionalities and statistics.}
\label{tabelka}
\end{center}
\end{table}
It is worth noting, that as was derived in \cite{Micheli2010}, collision rates for $y=1$ and without the application of the electric field have similar magnitude in different number of dimensions, provided equivalent gas density (cf. tab. \ref{tabelka}).
\begin{equation}
\label{uniw}
\mathcal{K}_j^{\mrm{re}}\approx \frac{4 \pi \hbar}{\mu}g_j L_j(\kappa)\frac{\rho}{a_h^{3-N}}
\end{equation}
where $j=0$ ($j=1$) stands for even (odd) partial waves, $g$ depends on dimensionality, $\frac{\rho}{a_h^{3-N}}$ is an equivalent three-dimensional gas density taking into account $a_h$ (squeezing in $3-N$ space dimensions), and $L_j(\kappa)$ is a parameter that has dimension of a length, and is defined in Table~\ref{tabelka}.

\section{Scattering length resonances and collision rates}

In this section we would like to present results of our numerical calculations. The multichannel wave function was propagated using the Numerov method, where at short range we imposed semiclassical boundary conditions Eq.~\eqref{Psi}, defined in terms of $s$ and $y$ parameters. We focus here on fermions, mostly for comparison with experimental results for KRb \cite{Miranda2011}.
In order to generalize the outcomes for a whole class of elements and molecules all quantities are expressed in dimensionless units. Besides the units defined before, we have also used the dipolar length $a_d=\mu d^2/(\hbar^2)$  ($\mu$ - a reduced mass, $d$ - an induced dipole moment) and a characterstic length associated with the trap $a_\mrm{h}=\sqrt{\hbar/(\mu \Omega)}$ ($\Omega$ - trap frequency). Moreover in the subsequent equations $q$ will denote a wave vector of colliding particles.

\begin{figure}
  \begin{center}
    \includegraphics[width=0.78\linewidth,clip]{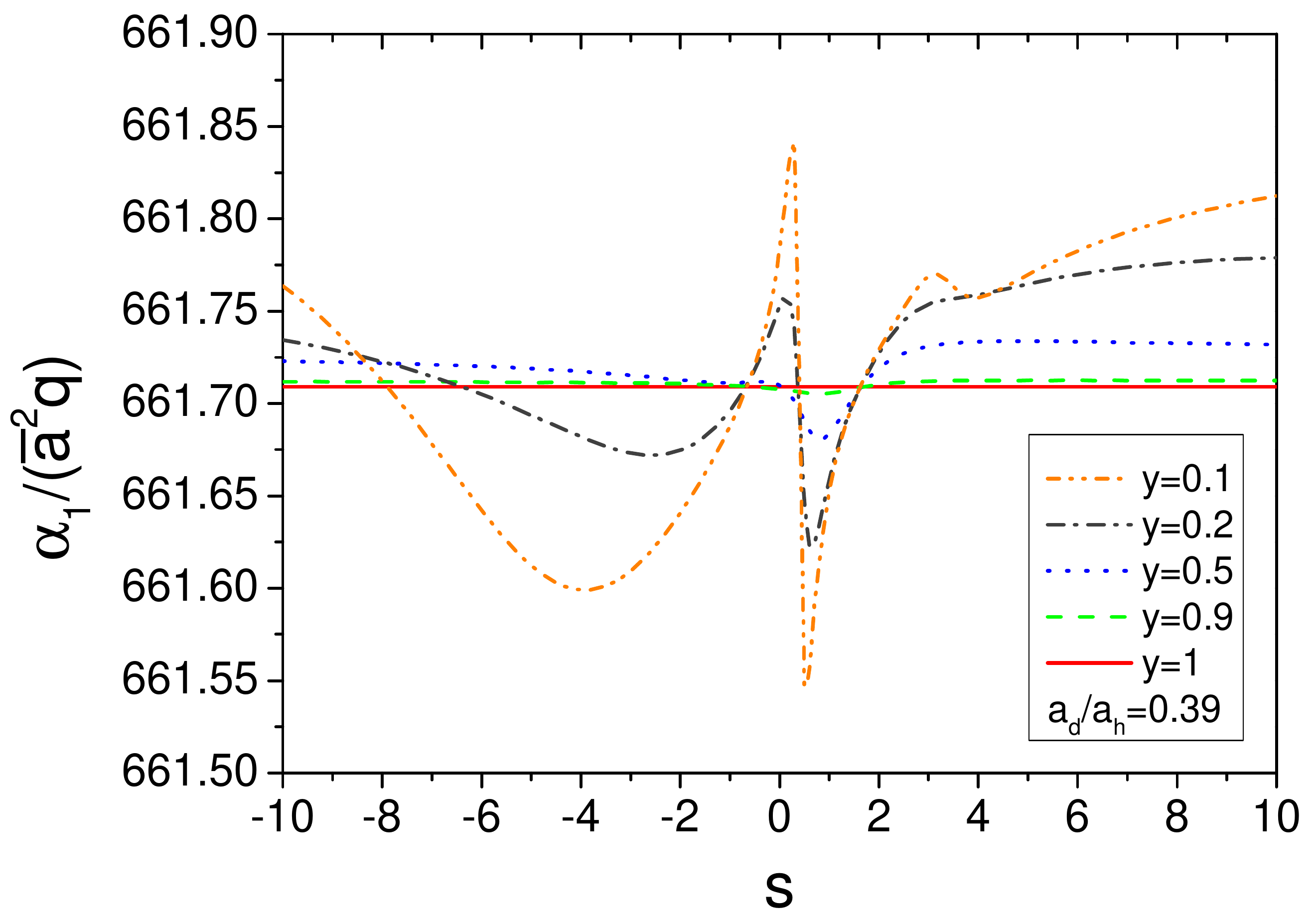}
  \end{center}
  \caption{The real part of the scattering length for different short-range reaction probabilities and the nonzero induced dipole moment for $q \bar{a}=0.05$.}
  \label{fig:6}
\end{figure}

\begin{figure}
  \begin{center}
    \includegraphics[width=0.78\linewidth,clip]{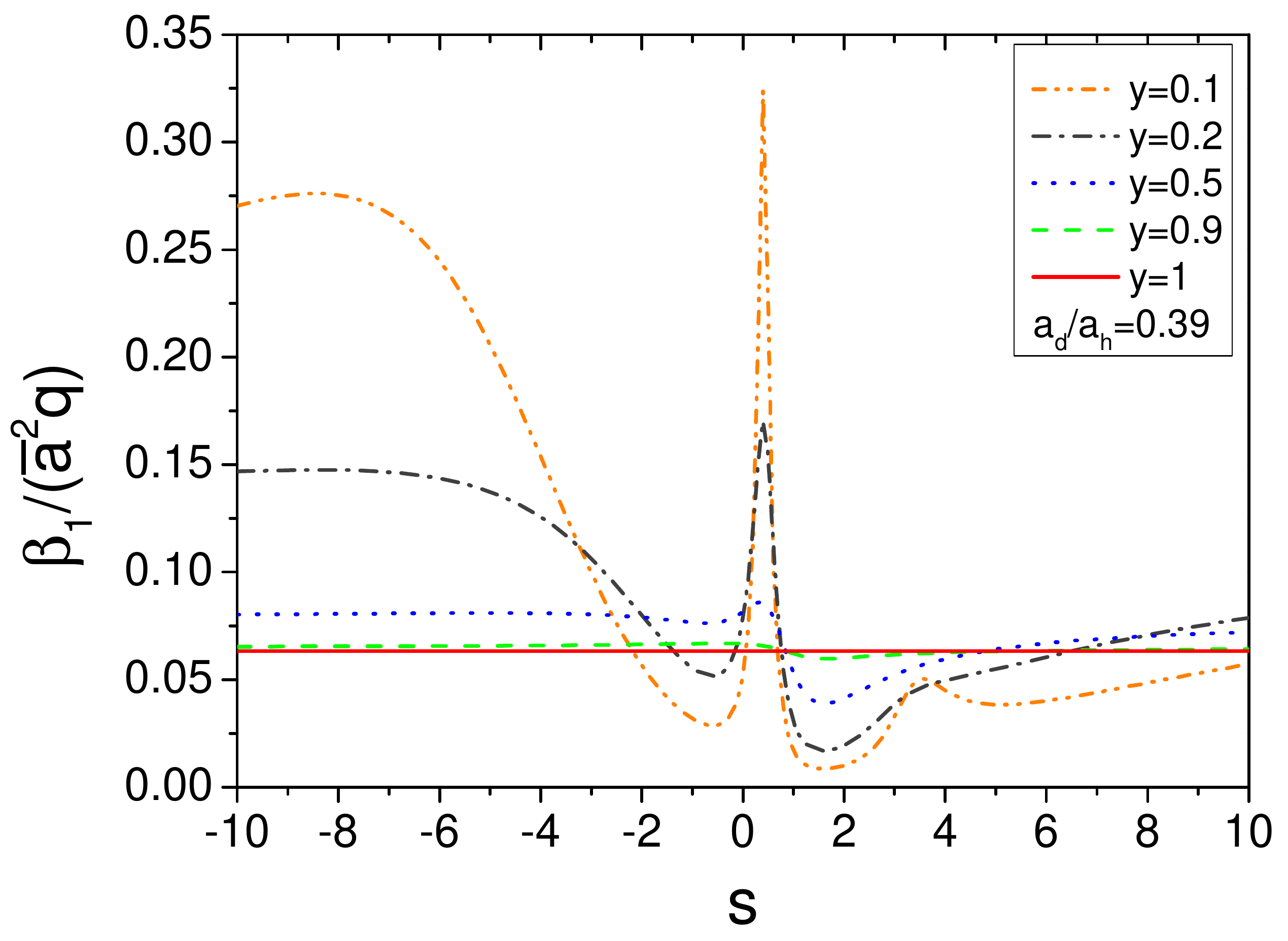}
  \end{center}
  \caption{The imaginary part of the scattering length for different short-range reaction probabilities and the nonzero induced dipole moment for $q \bar{a}=0.05$.}
  \label{fig:7}
\end{figure}

Let us start from the analysis of the complex scattering length (\ref{rozpraszania}). For dipolar length of the order of the trapping length, the amplitude of its real part is much larger than the amplitude of its imaginary part (it can be observed on Figs.~\ref{fig:6} and \ref{fig:7}). Hence, with help of Eq.~(\ref{szybkosci}) we can infer, that the real part of the scattering length for ultralow energies (here of the order of $q\bar{a}=0.05$, that would be equivalent to $50$nK for KRb) is responsible for the elastic scattering rate, whereas the imaginary part governs the reactive collisions.

Fig.~\ref{fig:6} depicts the real part of the scattering length for the fixed dipolar length and the trap length. We have investigated confinement-induced resonances. For the unit probability of reaction at small separation ($y=1$) resonances are not present, but with diminishing $y$ parameter (responsible for the reactivity and the outgoing probability flux) the amplitude of resonances increases.

A similar behavior can be observed for the imaginary part, which is shown in Fig.~\ref{fig:7}. It may be counterintuitive that with lowering the reaction probability at small distances, the reactive collision rate may increase close to the resonance. It can be easily explained. For small $y$ parameter the molecules can stay for a longer time at short distances and in case of their energy being in resonance (possible only for $y<1$) the collision rate can be highly enhanced. What is interesting, a resonance, that for pure van der Waals interaction occurs at $s=2$ for fermions, has been strongly shifted by the tight trapping potential and the applied electric field to $s\approx 0.4$.

\begin{figure}
  \begin{center}
    \includegraphics[width=\linewidth,clip]{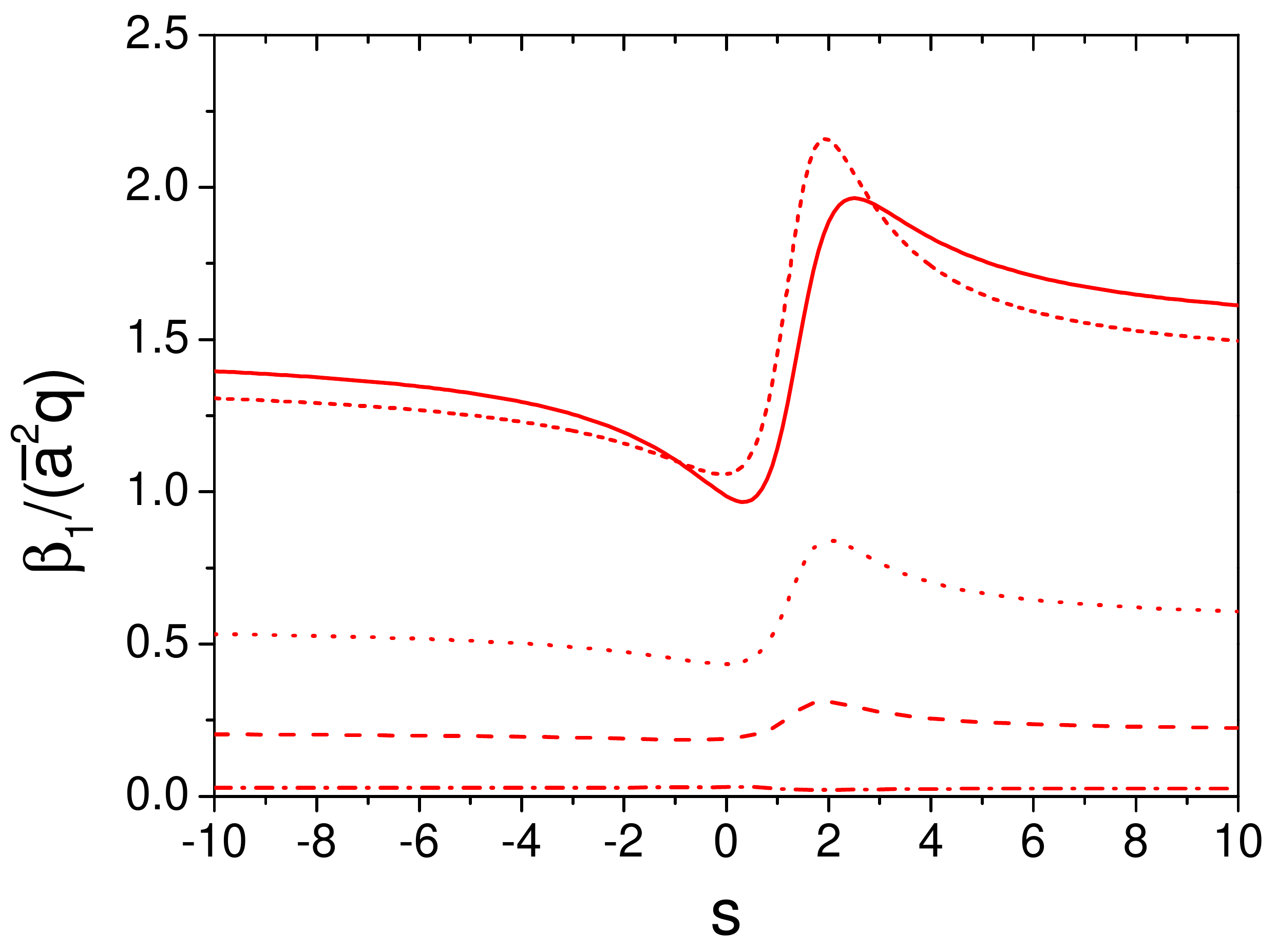}
    \includegraphics[width=\linewidth,clip]{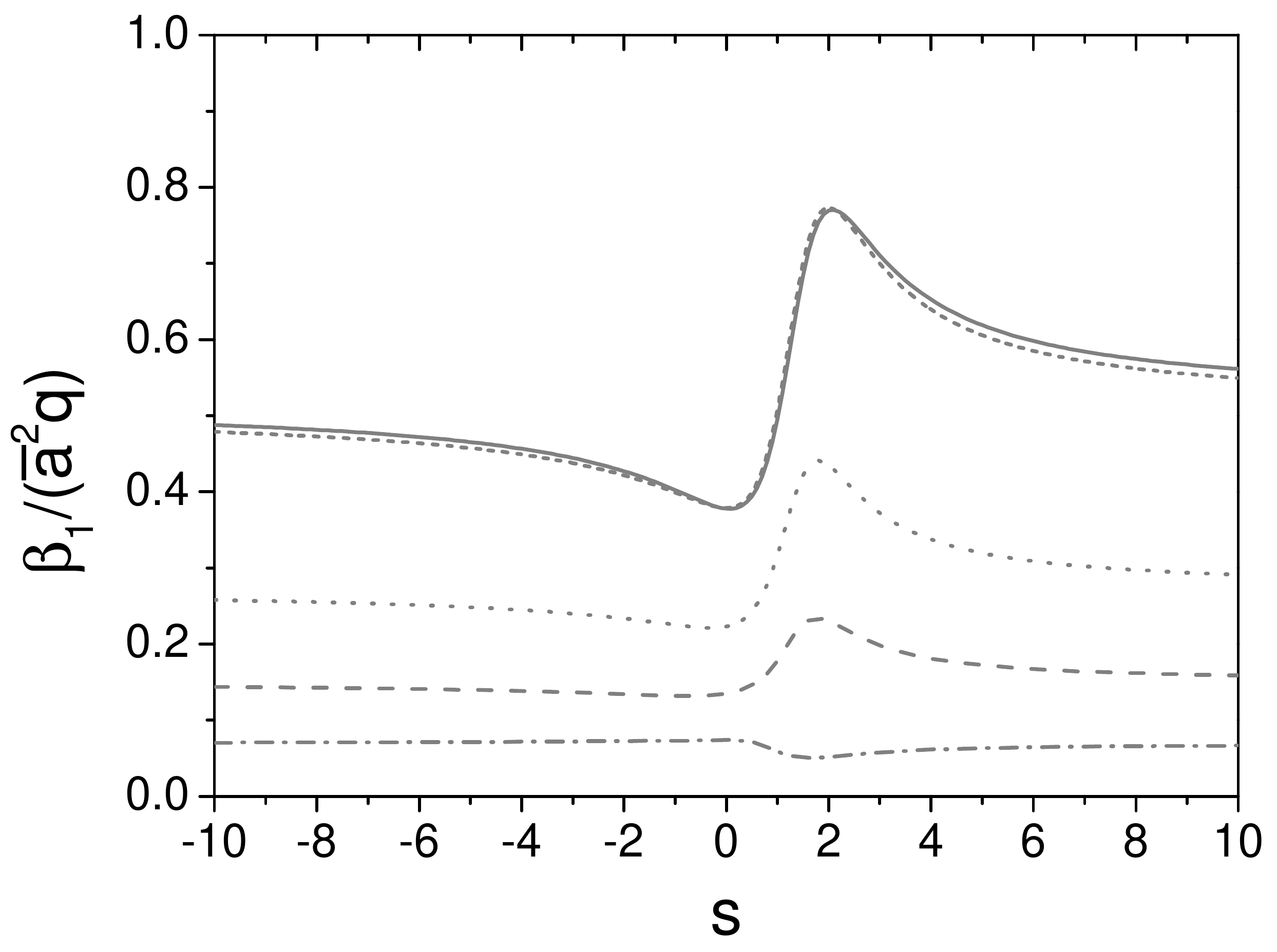}
  \end{center}
  \caption{Imaginary part of the complex scattering length for different induced dipole moments and the trapping strength. Upper panel:  $a_{h}/\bar{a}=1.7$, lower panel: $a_{h}/\bar{a}=5.2$, short dashes: analytical approximation (equation (42) in \cite{Idziaszek2015}) for $a_d/\bar{a}=0$, continuous line: numerical calculations for $a_d/\bar{a}=0$, dots: $a_d/\bar{a}=0.32$, dashes: $a_d/\bar{a}=0.73$, dots and dashes: $a_d/\bar{a}=2.02$. All calculations for $y=0.7$.}
  \label{fig:8}
\end{figure}

Let us compare the previous outcomes with Fig.~\ref{fig:8}. Here, the imaginary part of the scattering length is depicted as a function of $s$ parameter, but also for different traps and induced dipole moments. Firstly, we can observe, that the resonance is shifting not only with reactivity, as was seen before, but also with growing induced dipole moment towards lower $s$ values. Moreover, for the strong enough applied electric field, the resonance is almost totally damped. Similarly with growing squeezing in $z$ direction, the amplitude of the resonance gets lower and shifted (mind the scale on $y$ axis in Fig.~\ref{fig:8}). Hence, we were able to qualitatively verify, that for sufficiently squeezed induced dipoles the {\it head to tail} \cite{Miranda2011} collisions are being inhbited, because they are energetically not favourable and {\it side by side} collisions do dominate. The limitation of resonances may be partly understood by the fact that in {\it side by side} collisions the interaction potential is highly repulsive and molecules do not have a chance to meet at short distances and react. The outcomes without the electric field are compared with an analytic approximation (short dashes on Fig.~\ref{fig:8}) for a Dirac delta type interaction \cite{Idziaszek2015}. 

\begin{figure}
  \begin{center}
    \includegraphics[width=0.5\linewidth,clip]{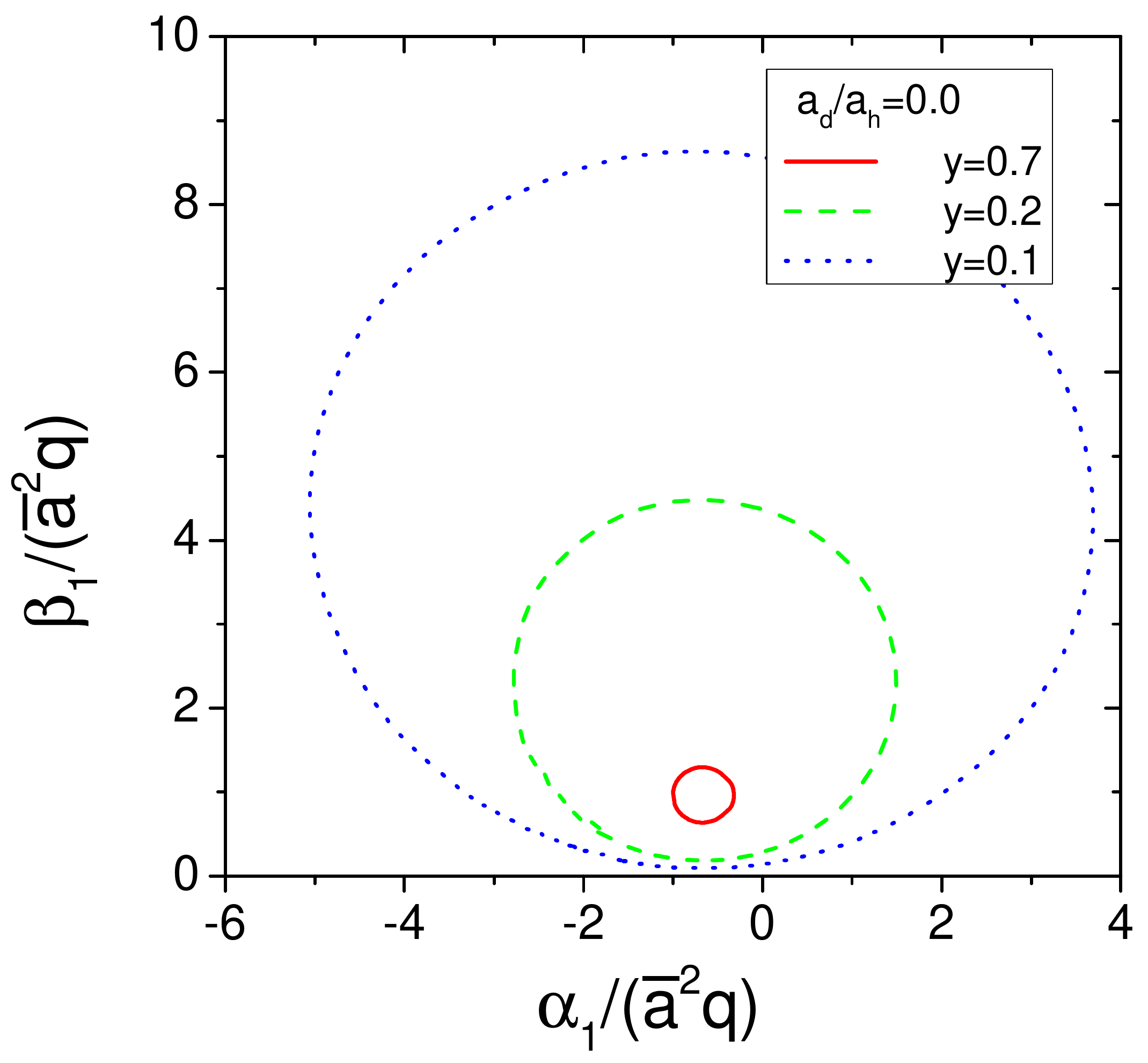}\includegraphics[width=0.5\linewidth,clip]{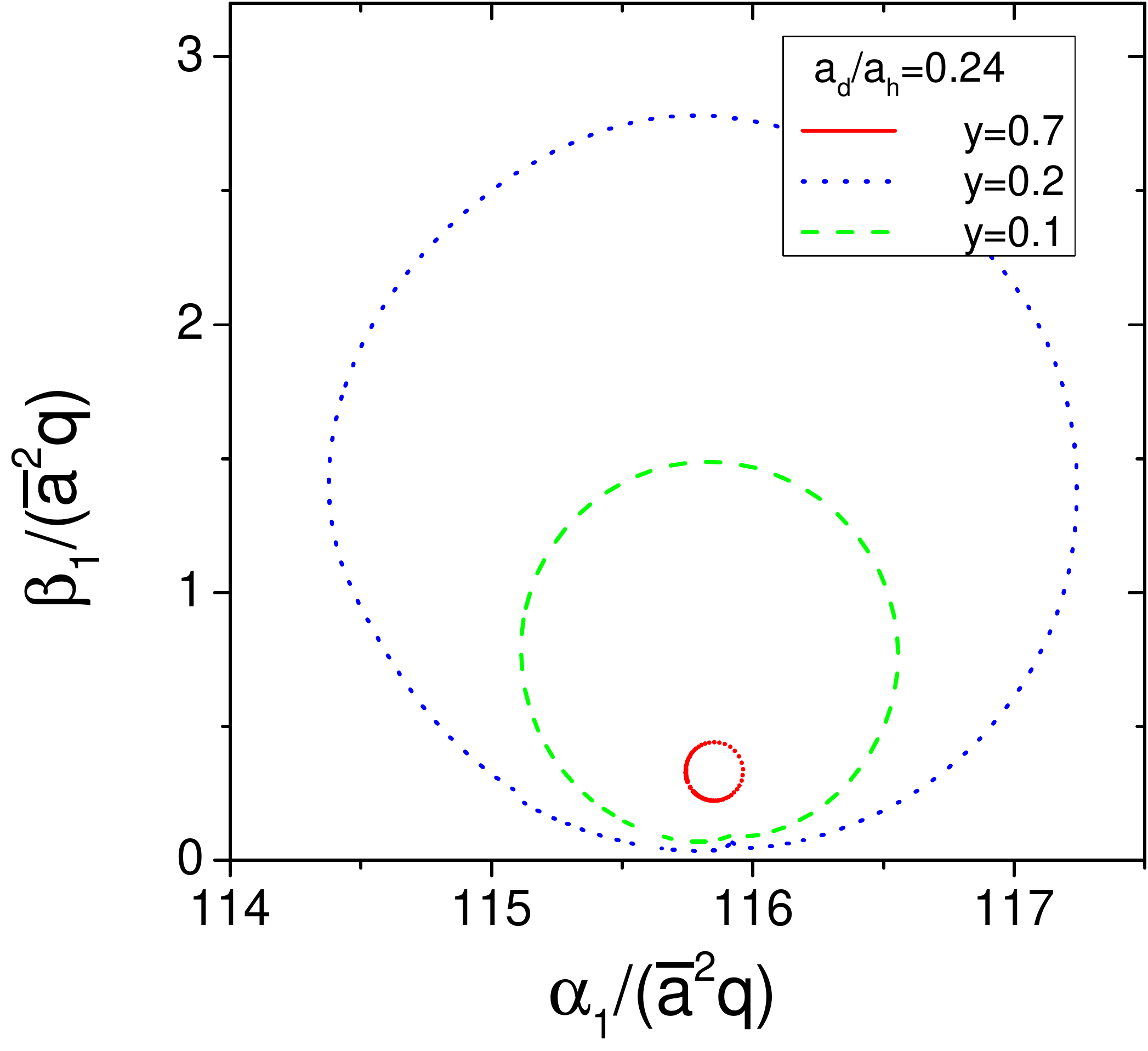}
    \includegraphics[width=0.5\linewidth,clip]{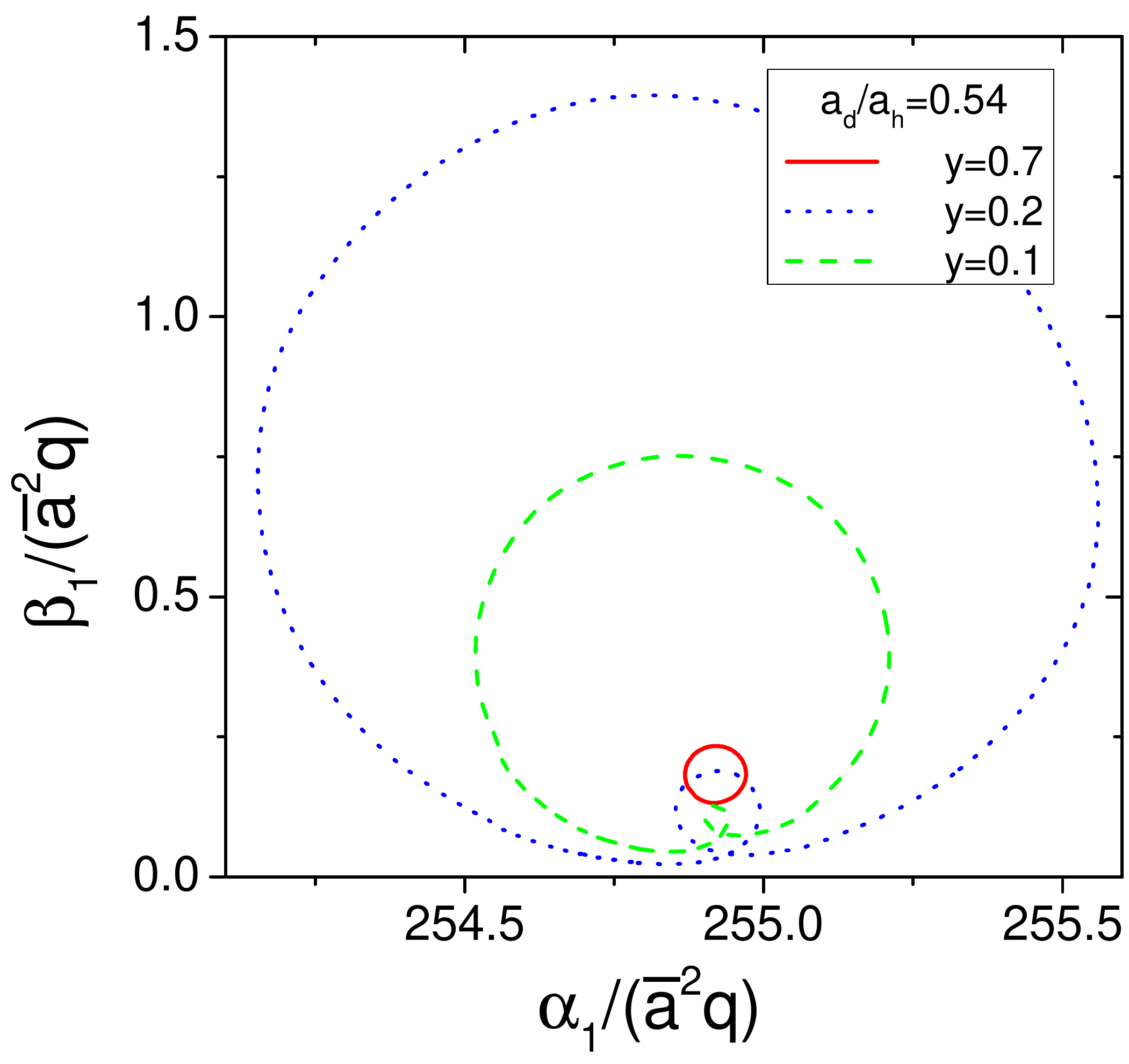}\includegraphics[width=0.5\linewidth,clip]{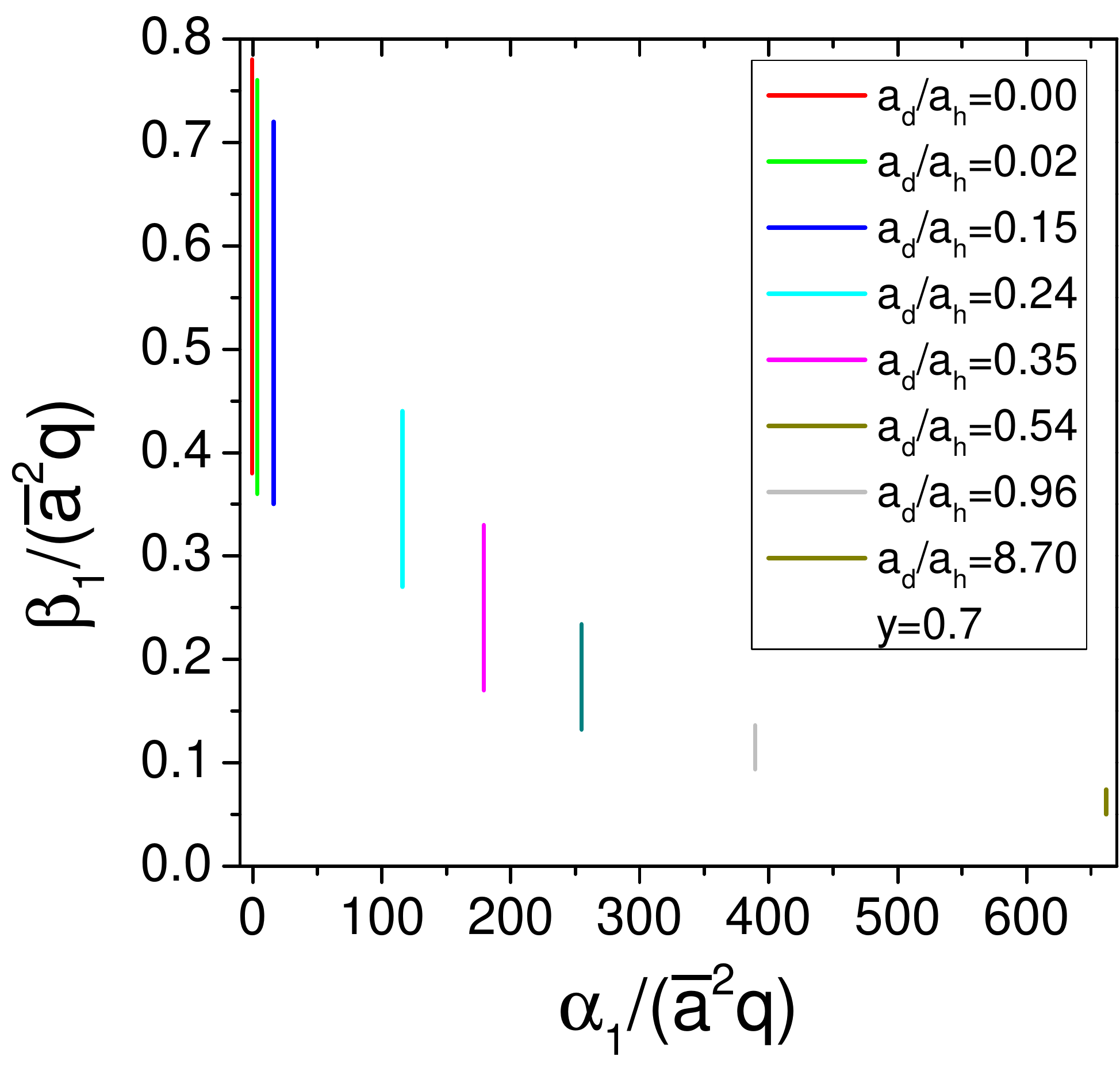}
  \end{center}
  \caption{Parametric plots of the imaginary and the real part of the scattering length for all possible values of $s$ parameter. Note a different scale on $x$ and $y$ axes in the bottom right panel.
  }
  \label{fig:Kolka2}
\end{figure}

In Fig.~\ref{fig:Kolka2} we show parametric plots of the real and imaginary parts of the scattering length as a function of the phase parameter $s$, for a fixed value of $y$. It turns out that for not too strong values of the dipole moments, they are circles with radius depending on reactivity at short distances.
For universal systems ($y=1$) the circle reduces to a single point, whereas with a diminishing reactivity the radius increases. It means that for different short-range phases (so also $s$ parameters) for small short-range reactivity large reactive and elastic rates may appear. 
Bigger dipolar length leads to shifting of the circles towards larger values of the real part of the scattering length and also smaller values of the imaginary part of the scattering length. Meanwhile the circle shrinks. We can conclude that with help of equation for reactive and elastic collision rates \eqref{szybkosci} even without the precise knowledge of $s$ parameter, that picks a certain point from the circle for a certain short-range reactivity it is easy to find a range of possible reactive and elastic collision rates. What is more for large applied electric fields the system approaches the universal limit immune to the short-range parameters.

\begin{figure}
  \begin{center}
    \includegraphics[width=0.74\linewidth,clip]{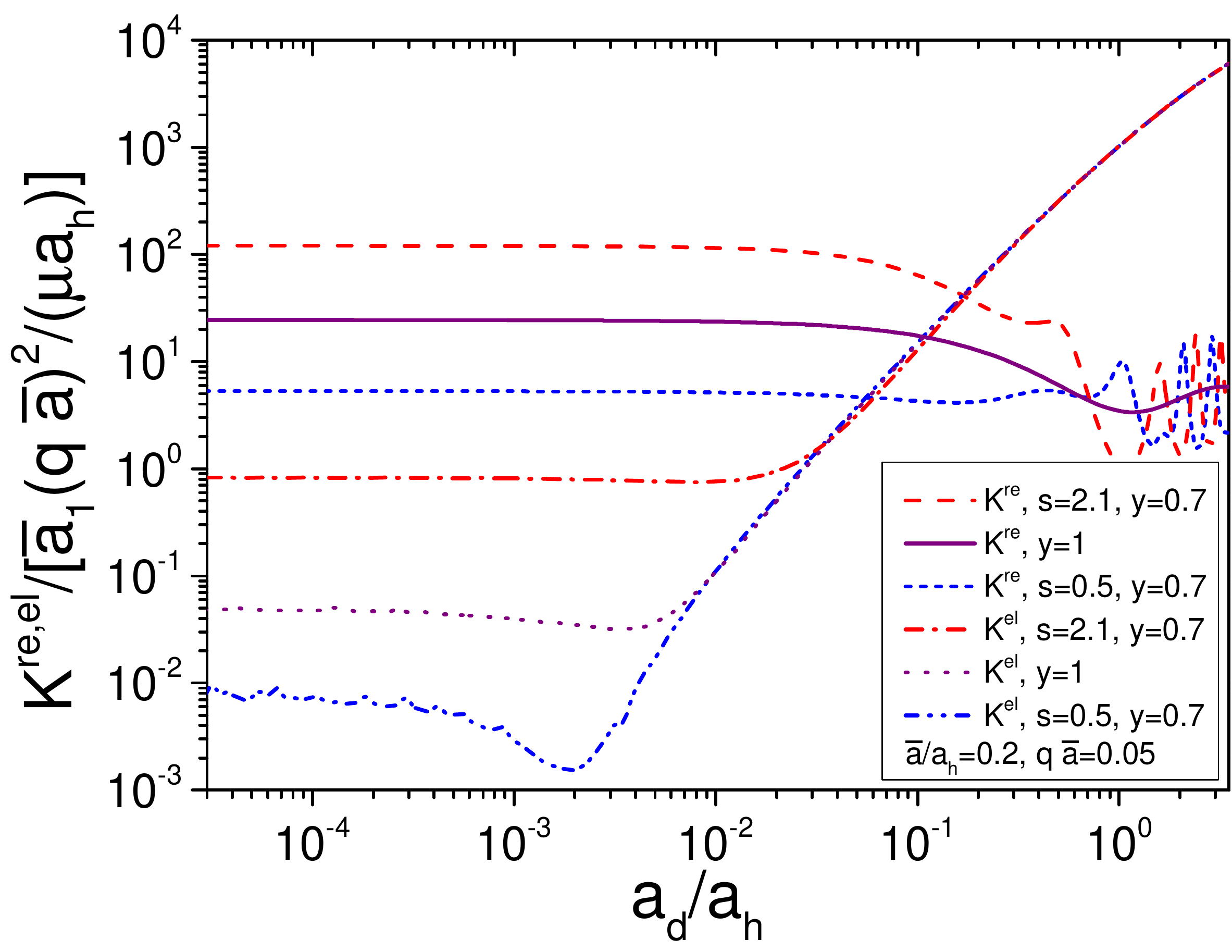}
  \end{center}
  \caption{Reactive and elastic collision rates as a function of the ratio of dipolar length to the characteristic trap length and the phase given by the $s$ parameter for intermediate values of experimental parameters.}
  \label{fig:s}
\end{figure}

On Fig.~\ref{fig:s} we have shown collision rates for different values of $s$ and $y$ parameters as a function of dipolar length. Elastic collisions despite differences in the scattering length, approach the universal limit ($y=1$) for sufficiently high external electric fields, so they become $s$ parameter independent. Similarly reactive collisions for sufficiently high dipolar length do oscillate around the universal curve. The oscillations that are present are the effect of shape resonances. So for strong dipoles elastic collisions dominate and it should be possible to stabilize the system and for example conduct the evaporative cooling.

\begin{figure}
  \begin{center}
    \includegraphics[width=0.74\linewidth,clip]{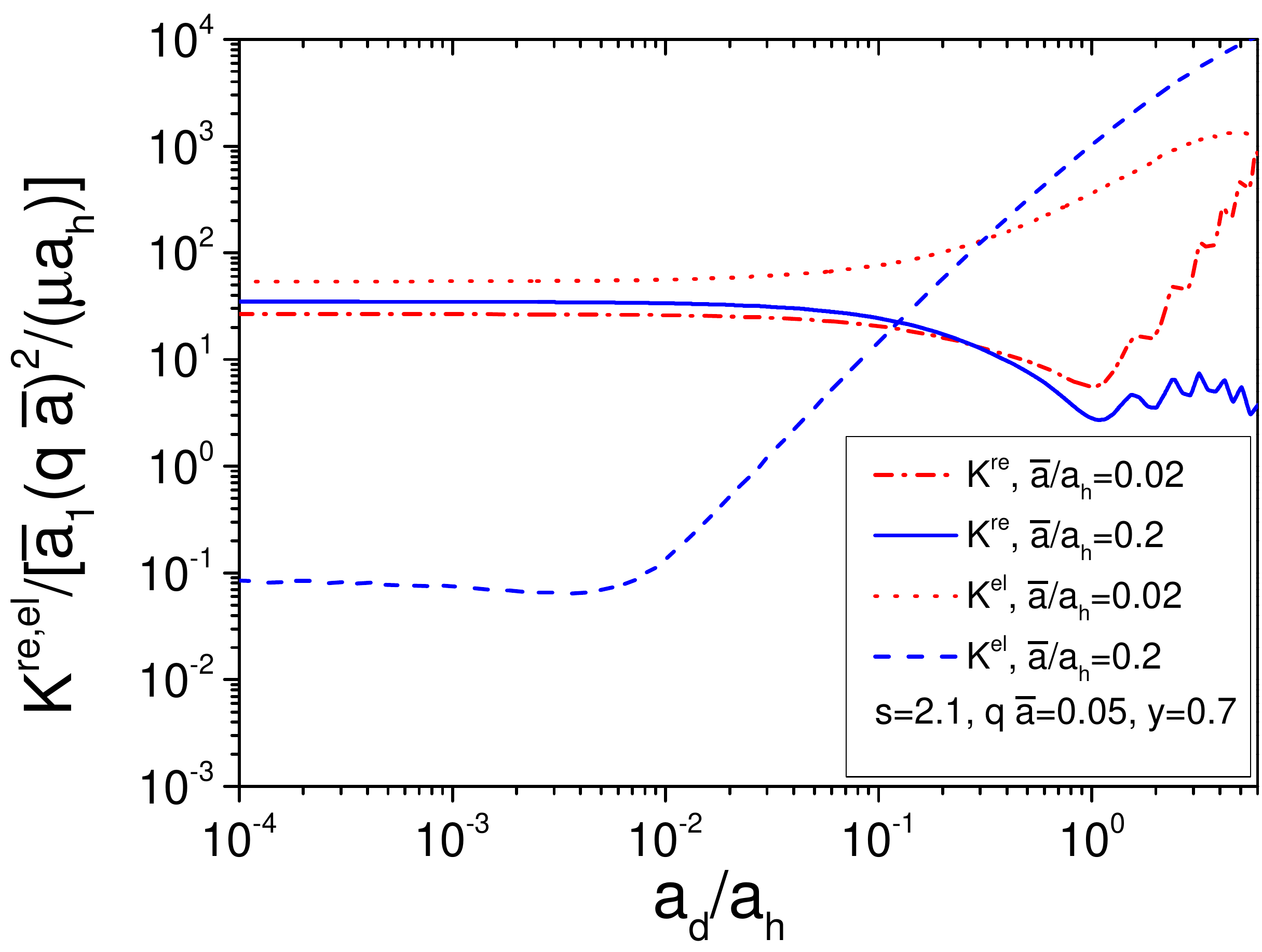}
  \end{center}
  \caption{Reactive and elastic collision rates as a function of the ratio of dipolar length to the characteristic trap length and the geometry of the trap for trap parameters corresponding to a very weak trap ($\bar{a}/a_\mrm{h}=0.02$, dash-dotted and dotted red lines for reactive and elastic collision rates, respectively) and for moderate trap parameters ($\bar{a}/a_\mrm{h}=0.2$, solid and dashed blue lines for reactive and elastic collision rates, respectively). The figure presents the influence of the trap dimensionality on the elastic and reactive rates.}
  \label{fig:scisniecie}
\end{figure}

 Now, we will look closer at the dimensionality of the system. We investigate how strong should be the dipole moment to observe a crossover from a 3D to the quasi-2D regime. Let us focus on Fig.~\ref{fig:scisniecie}. For a strong trapping (blue lines) and weak dipoles reactive collisions are dominating, but for larger dipolar lengths their amplitude is diminishing and elastic collisions win. This is completely another way for a weak trap with a large characteristic lengthscale (red lines). Here, reactive collisions are for small applied electric fields less favourable than elastic ones, but finally they aquire similar level. This growth is connected to the fact that there is a space for molecules to collide {\it head to tail}, that enhances reactivity. The theoretical outcomes presented in this paper are valid for a broad class of dipolar molecules. Similar conclusions may be inferred from experimental investigations~\cite{Miranda2011}, where the presence of the tight trapping potential in the plane perpendicular to the external electric field significantly supresses the reactive to elastic rate ratio.

\begin{figure}
  \begin{center}
    \includegraphics[width=0.74\linewidth,clip]{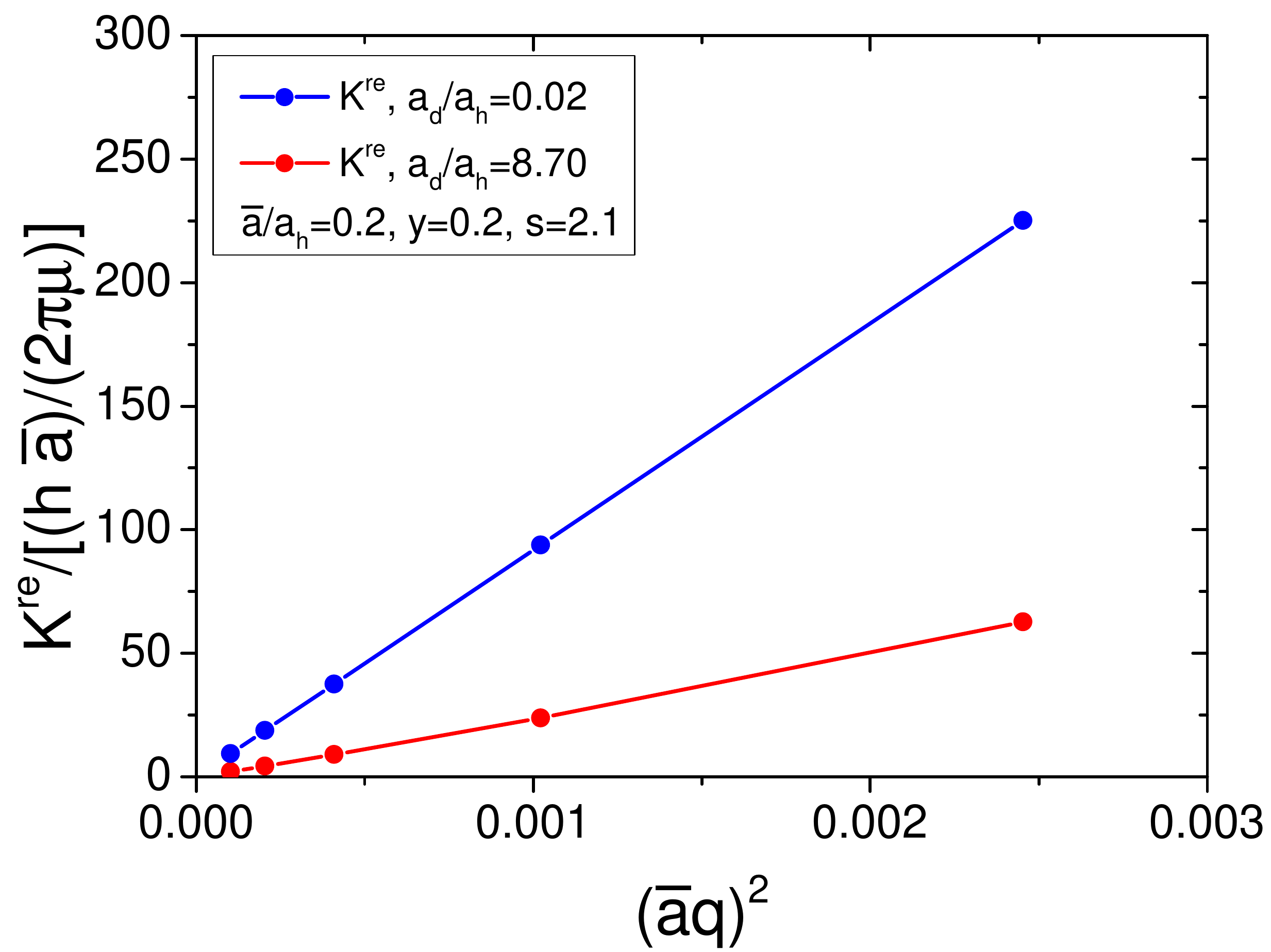}
  \end{center}
  \caption{Reactive collision rates as a function of the scattering kinetic energy. Linear dependence is clearly visible for nonzero dipole moments.}
  \label{fig:ztemperatura}
\end{figure}

 Finally we will check how reactive collision behave as a function of the kinetic energy of the colliding particles. It is depicted in an dimensionless units in Fig.~\ref{fig:ztemperatura}. For two different dipolar lengths the dependence is linear, in an agreement with a Wigner law for an inelastic collisions for $\ell=1$ and with theoretical predictions for universal collisions \cite{Li2008, Micheli2010}.

\section{Concluding remarks}

We have investigated reactive and elastic collisions in a quasi-2D geometry in the presence of an external electric field. We have studied the influence of the trapping and the induced dipole moment on the resonances of the complex scattering length and collision rates. We have shown that for sufficiently large induced dipole moments elastic collisions dominate over reactive ones and the system is stabilized. Furthermore, we have shown, that reactive collision rates scale linearly as a function of the scattering energy even for nonzero induced dipole moments. For moderate and small dipole moments parametric plots of real and imaginary parts of the scattering length for various short-range phases are circles with diameter diminishing with increasing dipole moment. This is similar to calculations without the external electric field \cite{Idziaszek2010} and shows that even without the knowledge of the short-range physics we could predict the possible ranges of elastic and reactive scattering rates and their ratio.


\section{Acknowledgments}
We thank K. Jachymski for stimulating discussions and reading and commenting on the manuscript. This work was supported by the Foundation for Polish Science International PhD Projects Programme co-financed by the EU European Regional Development Fund and National Science Centre (Poland) Grants No. DEC-2012/07/N/ST2/02879 and DEC-2011/01/B/ST2/02030).

\bibliography{bibliografiadoktorat0013}

\end{document}